\documentclass[pra,twocolumn,superscriptaddress,showpacs,floatfix]{revtex4}

\usepackage{graphics}           
\usepackage{bm}                 
\usepackage{amsmath}            
\usepackage{amsfonts}
\usepackage{amssymb}
\usepackage{latexsym}		

\begin{document}

\title{Bounds on entanglement in qudit subsystems}

\author{Vivien M. Kendon}
\email{Viv.Kendon@ic.ac.uk}
\affiliation{Optics Section, Blackett Laboratory, Imperial College,
London, SW7 2BW, United Kingdom.}

\author{Karol \.{Z}yczkowski}
\email{karol@cft.edu.pl}
\affiliation{Center for Theoretical Physics,
Polish Academy of Sciences, Al. Lotnik\'{o}w 32/44,
02-668 Warszawa, Poland}

\author{William J. Munro}
\affiliation{Hewlett-Packard Laboratories, Filton Road, Stoke Gifford,
Bristol, BS34 8QZ, United Kingdom.}

\date{March 18, 2002; revised May 30, 2002; further revised September 17, 2002}

\begin{abstract}
The entanglement in a pure state of $N$ qudits ($d$-dimensional
distinguishable quantum particles) can be characterized by specifying
how entangled its subsystems are. A generally mixed subsystem of $m$ qudits
is obtained by tracing over the other $N-m$ qudits. We examine the
entanglement in 
the space of mixed states 
of $m$ qudits. We show
that for a typical pure state of $N$ qudits, its subsystems
smaller than $N/3$ qudits will have a positive partial
transpose and hence are separable or bound entangled.
Additionally, our numerical results show that the probability
of finding entangled subsystems smaller than $N/3$ falls
exponentially in the dimension of the Hilbert space.
The bulk of pure state Hilbert space thus consists of highly
entangled states with multipartite entanglement encompassing
at least a third of the qudits in the pure state.
\end{abstract}



\maketitle


\section{Introduction}
\label{sec:intro}

Quantum information is a rapidly developing field exploiting the
peculiar entanglement properties of quantum states.
Its applications can be broadly divided into two general types:
quantum computation \cite{divincenzo95a,vedral98b}, and quantum
communication \cite{schumacher96a}.
Quantum communication (including quantum cryptography and quantum
teleportation) can be framed in terms of repeated use of pairs
of entangled qubits -- the entanglement properties of two qubits have
been well characterized, and a number of analytical measures of entanglement
are known
\cite{bennett96b,vedral97a,wootters97a,zyczkowski98a,horodecki01a,vidal01a}.
However, for multi-qubit systems, which are essential for quantum computation,
few entanglement measures can be calculated even for pure states.  Despite
these difficulties, arrays of qubits have been the focus of recent attention
\cite{wootters00a,oconnor00a,koashi00a,gunlycke01a,raussendorf01a,buzek01a},
though often only pairwise entanglement has been considered in such systems.
The more general question of entanglement in multi-qudit systems
($d$--dimensional quantum particles) is also important
since most real systems (e. g., atoms) have more than two states.

In this paper we tackle the question of entanglement in larger qudit
systems by deriving a bound on the size of typical entangled
mixed states that are subsystems of larger pure states.
Previous numerical work by some of us \cite{kendon01a}
suggested that qubit subsystems somewhat less than half the size of the pure
state, typically have no usable entanglement.  In other words,
since pure states are typically highly entangled, the entanglement
is distributed in multi-partite entanglement involving around half or more of
the qubits.  This work was limited by computational resources to small
numbers of qubits (less than 14).
Here we provide analytical bounds for any number of qubits, and also
extend the analysis to any dimension of qudits.

We emphasize that we are interested in average properties of all
possible pure states.
It is easy to construct states with entanglement properties that lie
outside our bounds.  For example, the 
so-called $W$ states (symmetric superposition of one qubit
in state $|1\rangle$ with the rest in state $|0\rangle$)
have only pairwise entanglement
\cite{koashi00a} for any value of $N$.  However, such states are of 
small measure and do not contribute significantly to the average properties of
pure states sampled from the whole of Hilbert space.

To further motivate our investigations, we note a connection to the
the question of the origin of the speedup in quantum computing.
Recently Jozsa and Linden in \cite{jozsa02a}
argue that for an exponential speedup over classical computation, the
state of a quantum computer using pure states will have multi-partite
entanglement encompassing an arbitrarily large fraction of the
total number of qubits in the computation.  However, they also point
out that the entanglement cannot be inferred to be the cause of the speedup,
rather, it is one of a number of properties of the typical states in
the Hilbert space of the quantum computer that are necessary for such
a speed up.  Our work confirms this property of typical states in
Hilbert space, and also provides an analytical lower bound
on the number of parties
in the multi-partite entanglement for a given size of Hilbert space.

The paper is organized as follows.  First, in Sec. \ref{sec:defs},
we introduce clear definitions of the system we are considering
and the entanglement measures we will use to analyze it.
Next in Sec. \ref{sec:bounds}, we derive the analytic bounds, followed by
comparison with numerical results in Sec. \ref{sec:numerical}.  We
finish with conclusions in Sec. {\ref{sec:conc}.

\section{Definitions and entanglement measures}
\label{sec:defs}

In a finite dimensional Hilbert space ${\cal H}$ there
exists a natural, unitarily invariant, Fubini--Study measure $\mu_{FS}$,
induced by the Haar measure on the unitary group.
A system of $N$ qudits has a Hilbert space of size $d^N$.
We will investigate entanglement in such systems
by sampling randomly from the set of all pure $N$--qudit states
according to the natural measure $\mu_{FS}$ and
calculating their average, which is to say, typical properties.
This will give us useful information about the majority of states in
Hilbert space. 
Having chosen a random pure state of $N$ qudits, we can divide it into
two subsystems by partial tracing.  A subsystem of size $m$ is obtained
by partial tracing over the other $N-m$ qudits.  
This partitioning could represent the
system of interest plus the environment, for example, or two different parts
of one system. 
From the mathematical point of view
one obtains in this way a certain probability measure $\mu_{N,m}$
in the space of mixed quantum states acting in a $d^m$ dimensional Hilbert
space induced by the natural measure $\mu_{FS}$
on the space of pure states in ${\cal H}_{d^N}$ 
\cite{zyczkowski00a}.
This is a useful way to sample randomly from mixed states, since unlike
$\mu_{FS}$ for pure states, there is no unique natural measure
for the space of mixed states.

\subsection{Entropy of subsystems}

Consider now a pure state quantum system described by 
a density operator $\rho = |\phi\rangle\langle\phi|$
acting in a composite Hilbert space
${\cal H}_{MK}={\cal H}_M\otimes{\cal H}_K$ of dimension $MK$. 
The reduced density matrices defined by the partial trace, 
$\rho_M:={\rm tr}_K \rho$ and $\rho_K:={\rm tr}_M \rho$
characterize both subsystems.
It is well-known that the von Neuman entropies of both 
subsystems are equal, $S_M=S(\rho_M)=-{\rm tr} \rho_M \ln \rho_M= 
S(\rho_K)=S_K$.
The value of $S_M$ shows how entangled the two subsystems are
with each other.  If $S_{M}=0$, there is no entanglement
and the composite state may be factorized,
$\rho=\rho_M \otimes \rho_K$, but this is not a
typical case. 
The mean entropy of a subsystem 
averaged over the natural measure $\mu_{FS}$ 
is given by
\begin{equation}
\langle S_M\rangle = \sum_{j=K+1}^{MK} \frac{1}{j} \;
                            - \frac{M-1}{2K}
                   \simeq \ln M - \frac{M}{2K},
\label{eq:sent}
\end{equation}
where $M\le K$ and the approximation holds for $0 \ll M < K$.
This result was first conjectured by 
Page \cite{page93a} and later proved in \cite{foong94a,sen96a}.

In our case the subsystems
consist of $m$ and $N-m$ qudits, so  
substituting  $M=d^m$ and $K=d^{N-m}$
into (\ref{eq:sent}) we obtain the mean 
entropy $\langle S\rangle_{N,m}$ of the subsystem analyzed, where we
are now using the subscripts to remind us of the number of qudits in the
pure state ($N$) as well as the number in the subsystem ($m$).
Equation (\ref{eq:sent}) tells us that on average,
the smaller subsystem has nearly maximal entropy,
showing that the two subsystems are highly entangled with each other,
a typical pure state is highly entangled, see also \cite{kempe01a}.
The particular value of $S_{N,m}$ provides a good measure of entanglement
between the two subsystems, but says little about entanglement between the
qudits within a single subsystem.
Asymptotically, for ($N\rightarrow\infty$), $m\le N/2$ will
tend to be maximally mixed and therefore separable, but
to obtain useful results for finite $N$,
we need further information on the size of the region around the maximally
mixed state that contains separable states.

\subsection{Purity and mixedness}
\label{ssec:purity}

Next we will characterize how mixed 
a typical $m$--qudit state $\rho_m$ might be.
A convenient measure is the \textit{purity} 
\cite{weaire77a} defined by
\begin{equation}
r(\rho_m):= {\rm tr}(\rho_m^{2}).
\end{equation}
Pure states are defined to have $r(\rho_m) = 1$
while mixed states have $1/d^m \le r(\rho_m) < 1$. 
We will also refer to
$R:=1/{\rm tr}(\rho_m^{2}) \equiv 1/R$ as the
\textit{inverse participation ratio} (IPR)
of the mixed state $\rho_m$. 
The IPR thus ranges from 1 (pure) to $d^m$ (maximally mixed),
and larger IPR means the states are more mixed,
while larger purity means the states are more nearly pure.

Consider random states drawn according to the natural
measure $\mu_{FS}$ on the space of pure states in a
$MK$ dimensional Hilbert space.
As before, we apply a partial trace to obtain the reduced density
matrices $\rho_M$ and $\rho_K$.
In refs.  \cite{lubkin78a,zanardi00a,zyczkowski00a}
it was shown that the average purity of $\rho_M$ and $\rho_K$ is
equal to 
\begin{equation}
\langle r\rangle=\frac{M+K}{MK+1}
\label{eq:purity1}
\end{equation}
This result does not depend on the particular way
in which the $MK$--dimensional Hilbert space is decomposed,
but only on the initial and the final dimensionality of the spaces.
Thus it also holds in the problem we are analyzing here of
pure states of $N$--qudits reduced to $m$--qudits by partial tracing.
Substituting $M=d^{m}$ and $K=d^{N-m}$
the average purity of the system reads
\begin{equation}
\langle r\rangle_{N,m}=\frac{d^{m}+d^{N-m}}{d^{N}+1}
\label{eq:ipr}
\end{equation}
In other words, the averaging has been performed with respect to the
induced measure $\mu_{N,m}$.

Numerical investigation of the induced measures shows \cite{zyczkowski00a}
that for $m \lesssim N/2$ the probability distributions $P(r)$ 
are concentrated close to the mean value 
$\langle r\rangle$.   
 \begin{figure}
    \begin{minipage}{\columnwidth}
	\begin{center}
	    \resizebox{\columnwidth}{!}{\includegraphics{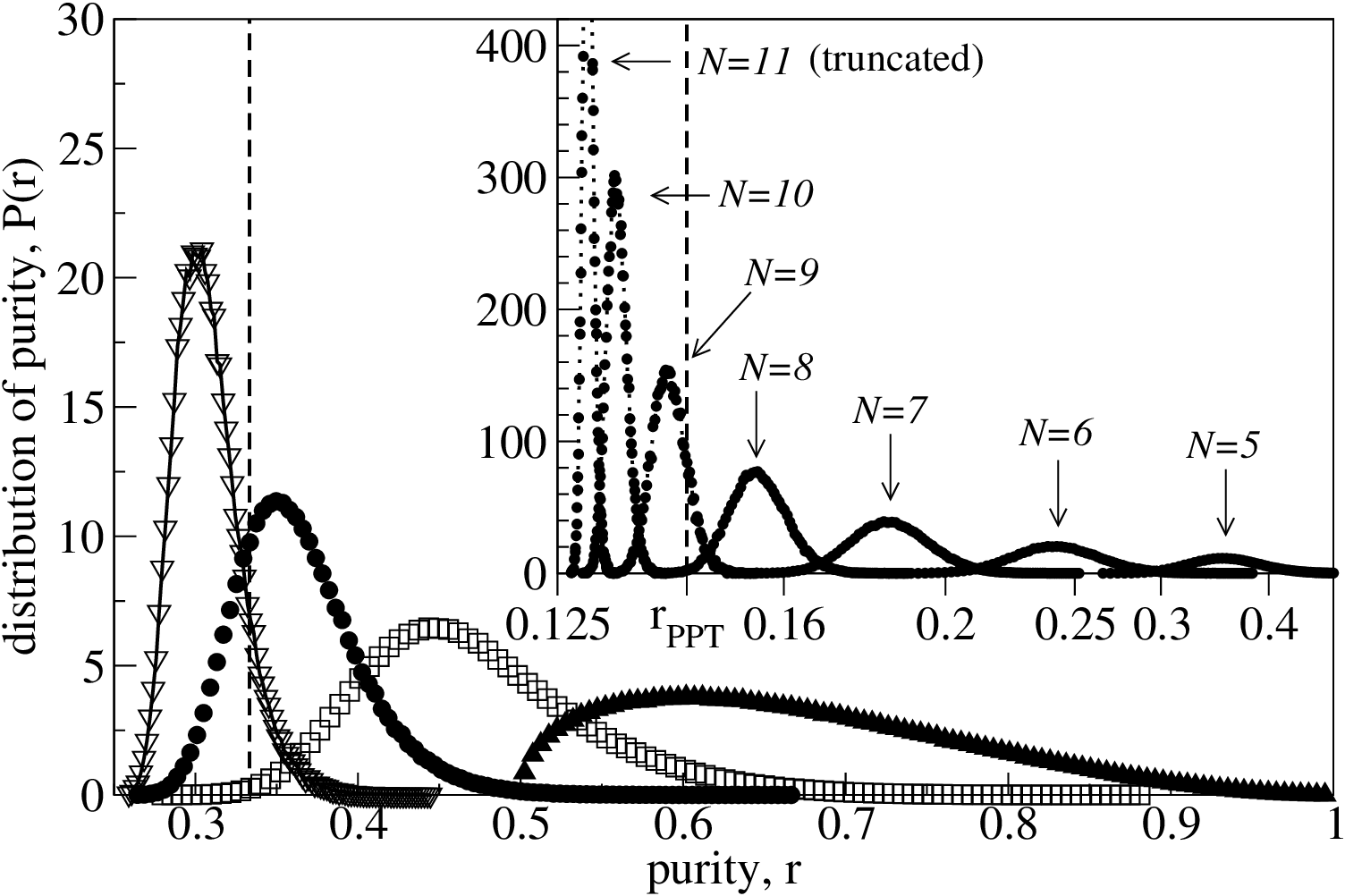}}
	\end{center}
    \end{minipage}
    \caption{The numerically obtained distribution of the purity $r$ for
	(main graph) mixed states of $m=2$ qubits obtained by partial tracing
	of random pure states of (right to left) $N=3$, 4, 5 and 6 qubits,
	and (inset) mixed states of $m=3$ qubits for $N=5$ to $10$.
	The vertical dashed lines indicate the values of $r_{\text{PPT}}$
	from eq. (\ref{eq:ipr-ppt}).}
	\label{fig:purdis2m}
 \end{figure}
This is illustrated in fig. \ref{fig:purdis2m}, where $P(r)$ is plotted for
$d=m=2$ and $N=3$ to 6 in the main plot, and for $d=2$, $m=3$, $N=5$ to 10
in the inset.  The scale of the purity axis on the inset has been adjusted
(the scale is in fact linear in $R=1/r$)
to show the data near the minimum purity more clearly.

Fixing the number $m$ of the qudits in the final system
and varying the initial number $N$ of qudits involved in the
pure state we get a sequence of probability measures
which sweep the space of the mixed states close to the
manifold of a constant purity.  
Hence it is justified to use the mean value $\langle r \rangle $
as a good estimate of the actual value $r$ for a typical subsystem.

\subsection{Peres criterion, positive partial transpose}

We will now describe a simple test for entanglement in multiparticle
mixed states due to Peres \cite{peres96a}.  
Consider a multiparticle mixed state $\rho$
acting in a $d^m$ dimensional Hilbert space,
which describes the system consisting of $m$ qudits.
An $m$--particle system can be split into
two parts containing $j$ and $m-j$ particles
respectively.  The partial transpose is applied to the $j$ particle 
part of the system producing $\rho^{T_j}:=({\mathbf 1}_L \otimes T_j)
\rho$. Here $T_j$ denotes the transpose operation in the $d^j$
dimensional
subspace, while the dimension of the subspace that has not been
transposed is
$L=d^{m-j}$. We say that the $m$--qudit state 
has the PPT property, 
if $\rho^{T_j}$ is positive for  all possible values of $j$
(it is sufficient to take $1 \le j \le m/2$) and
 choices of $j$ particles,
i. e., with respect to {\sl all} of its possible splitting into
two subsystems.
Conversely, a state which fails this test is described as being NPT
(negative partial transpose).  NPT states have some useful entanglement
that could be distilled with some probability into one or more
maximally entangled states of two qubits.
This test does not distinguish bound entangled states
\cite{horodecki97a,horodecky98a} from separable states,
but since bound entanglement is relatively rare \cite{zyczkowski99a},
and for most purposes it is the free entanglement that is 
useful, this still provides a useful characterization of entanglement.

For bipartite systems with Hilbert space dimension $d^2$,
it has been shown \cite{zyczkowski98a} that all the states
sufficiently close to the maximally mixed state,
with $R\ge d^2-1$ have a positive partial transpose (PPT).
This condition defines the maximal
ball inscribed in the convex body of mixed states.
Very recently it was shown that all mixed states
of a bipartite system 
belonging to the maximal ball posses not only the PPT
property but are also separable \cite{GurvitsBarnum02}.
The former result may be generalized for a multipartite case:
any state  $\rho_m$ 
(within the $d^m$ dimensional Hilbert space) for which
\begin{equation}
R\geq R_{\text{PPT}}=d^m-1
\quad \text{so that} \quad
r\leq r_{\text{PPT}}=\frac{1}{d^m-1},
\label{eq:ipr-ppt}
\end{equation}
has the PPT property with respect to any possible 
transpositions of the subsystems.
We give the proof of the generalization in Appendix \ref{sec:app_ppt}.

\section{Bounds on entanglement in subsystems}
\label{sec:bounds}

We are now ready to provide an analytical relationship between the
size of a subsystem and its typical entanglement properties.

\subsection{Bound on PPT region}

Using the fact that the distributions of the purity $P(r)$ are narrow
for the regions of interest ($m\lesssim N/2$), and thus
$\langle r\rangle_{N,m}$ is a good estimate of the actual value of
$r$ in a typical subsystem,
we apply this to the critical value $r_{\text{PPT}}$ obtained in the
previous section.
If the mean value $\langle r \rangle $
averaged over a certain induced measure is smaller that the critical value
$r_{\text{PPT}}$,  (or $\langle R\rangle > R_{\text{PPT}}$),
the majority of random states distributed with respect
to this measure are localized inside the maximal ball and are PPT.
We therefore combine eqs. (\ref{eq:ipr-ppt}) and (\ref{eq:ipr})
to calculate a relationship between $N$ and $m$
such that we expect almost all subsystems of $m$ qudits
will have a positive partial transpose.
We find $\langle r\rangle_{N,m} \le r_{\text{PPT}}$ when
\begin{equation}
d^{N} \ge d^{3m}(1-d^{-m}-d^{-2m}).
\end{equation}
Taking the logarithm (base $d$) gives us
an estimate of how many qudits the initial pure state
should contain, such that after the reduction to $m$--qudits
the obtained mixed state is PPT on average,
\begin{equation}
N_{\text{PPT}} \ge 3m+\log_{d}(1-d^{-m}-d^{-2m}).
\label{pptupex}
\end{equation}
The $\log_d$ term is always negative, becoming rapidly smaller for
either $m>2$ or $d>2$. Hence for systems initially
consisting of
\begin{equation}
N \geq 3m
\label{eq:upper}
\end{equation}
qudits
we expect the subsystems of size $m$ to be PPT
mixed states with a considerable probability.
For the simplest case  of $m=d=2$,
the smallest integer number larger than the right hand side of
(\ref{pptupex}) gives
$N_{\text{PPT}} = {\rm int}[6- \log_2(16/11)]+1 = 6$.
For a fixed  number  $N$ this
establishes a bound on $m$ below which subsystems
of this size have on average
no useful entanglement in an $N$--qudit pure state.
This is the main result in this paper.

It is also possible to calculate the maximum value of the
von Neuman entropy for states on the surface of the maximal ball,
see Appendix \ref{sec:app_vNent}.  This value $S_c$, given in
eq. (\ref{eq:entcrit}), can then be combined with eq. (\ref{eq:sent})
for the average entropy of a subsystem, to obtain a similar
relationship between $N$ and $m$.  The result is (as it should be)
the same as  eq. (\ref{eq:upper}), but with the restriction
$m\gg 0$ from the approximate form of eq. (\ref{eq:sent}).
Our derivation as presented using eqs. (\ref{eq:ipr-ppt}) and (\ref{eq:ipr})
is applicable to all $0<m<N$.

\subsection{Estimation of the transition region}

The above result is a bound on the average properties of typical
states that identifies the range of parameter values ($N\ge3m$) 
where there is a high probability that a subsystem of size $m$
is PPT.  We would also like to characterize the range of $N,m$ for which with
high probability the subsystems of size $m$ are entangled.
This region is not simply the inverse of the PPT region
because while the criterion we are using in eq.
(\ref{eq:ipr-ppt}) is sufficient for ensuring the subsystem is PPT, it
is not a necessary condition.  There are many states which are PPT that
have $r > r_{\text{PPT}}$ right up to separable pure states with $r=1$.

Since we do not, in general, know the size of the region of PPT
states outside the maximal ball defined by eq. (\ref{eq:ipr-ppt}),
our approach is to identify a region of the body of mixed states 
of a considerable size, which contains entangled states only.
There is a one parameter family of
mixed states (known as the generalized Werner states)
containing $m$ qudits defined as a mixture of the maximally
entangled pure state $|\Psi\rangle$ and the maximally mixed state $I_{d^m}$,
\begin{equation}
\rho_{\varepsilon}=\frac{(1-\varepsilon)}
{d^m} I_{d^m} +\varepsilon|\Psi\rangle\langle\Psi|
\end{equation}
where
\begin{equation}
|\Psi\rangle=\frac{1}{\sqrt{d}}\sum_{i=1}^d |i\rangle_1
\otimes \ldots \otimes |i\rangle_m
\end{equation}
and $\varepsilon$ is a real parameter between $0$ and $1$
specifying the proportions of the mixture.
These states have been shown to be entangled for
\cite{deuar00a,pittenger00a,rungta01a}
\begin{equation}
\varepsilon>\varepsilon_{ent}= \frac{1}{d^{m-1}+1}.
\end{equation}
and strictly separable for $\varepsilon\leq\varepsilon_{ent}$ (there are
no bound entangled states in this family).
This boundary at $\varepsilon_{ent}$ defines a set of entangled
states of positive measure \footnote{
By ``positive measure'' we mean that we have divided the Hilbert space
into two regions with measures of comparable size in $\mu_{N,m}$,
i. e., the ratio of the measures of the two parts is finite
(neither infinite nor zero).
} as can be seen as follows.
The states $\rho_{\varepsilon}$ lie on a line that is on an axis of rotational
symmetry in the set of mixed states.
This line intersects the set of separable states at
$\varepsilon = \varepsilon_{ent}$.
The set of separable states is convex, so
beyond the hyperplane normal
to the $\rho_{\varepsilon}$ line
there exists a set of entangled states of a positive measure 
(see also \cite{lockhart00a}).

It is straight forward to show that 
the purity for $\rho_{\varepsilon}$ at this boundary point
($\varepsilon = \varepsilon_{ent}$) is given by
\begin{equation}
r_{ent}=\frac{d^{m}+d^{2}+2d}{\left(d^{m}+d\right)^2}.
\label{eq:ipr_ent}
\end{equation}
Thus there exists a greater-than-zero probability to encounter
entangled states with $r > r_{ent}$.
However, we don't know that entangled states actually dominate over PPT
states for this value of $r$.  Since Werner states are among the most
entangled states \cite{munro01a,ishizaki00a,verstraete01a}
for a given purity $r$, we might actually guess that this
estimate will turn out to be too tight. 
Nonetheless, it provides a useful independent check on our previously
derived bound, so we present it anyway.

Equating $\langle r\rangle_{N,m}$ from eq. (\ref{eq:ipr})
with $r_{ent}$ from (\ref{eq:ipr_ent}) allows us to establish
an estimate for the entangled side of the transition between
entangled and PPT states.
 It is easily shown that $\langle r\rangle_{N,m} \ge r_{ent}$ when
\begin{equation}
N_{ent} \le 3m-2 + \log_{d}[1+(2d+1)d^{-m}+(d+2)d^{1-2m}].
\label{eq:ext}
\end{equation}
For all $d$ and $m$, the $\log_d$ term is positive and tending to zero
for increasing $d$ and $m$.  Hence for all finite $N$ with
\begin{equation}
N \leq 3m-2
\label{N3m2}
\end{equation}
we expect  the probability of finding the subsystem of size $m$
entangled (not PPT), to be non-zero.
Looking at the smallest case \footnote{
For 2 qubits, $\varepsilon_{ent}$ is at the boundary of the maximal ball
(PPT implies separable in this case, see also \cite{GurvitsBarnum02}),
which is why the exact results for the bounds for two qubits are
the same (r.h.s. of eqs. (\ref{eq:upper}) and (\ref{eq:ext}) are equal).  
But this isn't so for $d>2$ or $m>2$, hence
we get a different value for the bound from this method in general.
The states in between are maximally entangled mixed states,
see, \cite{munro01a,ishizaki00a,verstraete01a}.
}, $d=m=2$, we have
$N_{ent}={\rm int} [4 + {\rm log}_2(11/4)] = 5$.
Let us emphasize again, that this is an estimation only;
for this (or lower) values of $N$ the probability of
finding entangled subsystems of size $m$ is non-zero,
but it does not rule out the existence of substantial 
numbers of PPT subsystems of size $m$, so the transition might occur
effectively for even smaller values of $N$.

Now let us fix the initial size of the pure states of $N$ qudits
while we decrease the size of the final system of $m$ qudits.
For $m=N$ the probability of finding a separable pure state
is equal to zero. As $m$ is reduced, the states obtained
by partial tracing over $N-m$ qudits become increasingly mixed,
and the probability of finding PPT states increases.
Putting both results together we may characterize
quantitatively a transition region
\begin{equation}
 3m - 2 \lesssim  N \le 3m,
\label{eq:twobound}
\end{equation}
which does not depend on the qudit size $d$,
in which we estimate that PPT subsystems come to dominate over 
entangled subsystems.
(Note that the inequalities have reversed because we are identifying
the region where the ratio of PPT to NPT (negative partial transpose)
subsystems of size $m$ is of
order unity, rather than the regions where one case dominates over the other.)
These are surprisingly tight estimates
 of the boundary between entangled and
PPT states in terms of the number
of qudits in the state.  However, it
must be remembered that the number
 of qudits is a logarithmic measure of
the size of the Hilbert space. The addition of one qudit enlarges
the Hilbert space by a factor of $d$.

\section{Numerical comparisons}
\label{sec:numerical}

In \cite{kendon01a}, for the qubit
case $d=2$, it was shown numerically
that entanglement in a mixed $m$--qubit subsystem
of a typical $N$--qubit pure state
falls off sharply towards zero with increasing $N$ for
values of $m \le 5$ and $N \le 13$.
These numerical results suggest the transition from NPT states
dominating to PPT dominanting
is complete by $N\simeq 2m+3$, which falls outside the range of
eq. (\ref{eq:twobound}) for $m > 4$.  This is at the edge of
the results presented in \cite{kendon01a},
so we ran further numerical studies to confirm this
divergence between numerical and analytical results
and also, since our analytical results apply for any dimension
of the constituent quantum particles $d$, to test qutrits
and confirm they follow the same pattern.

Using straightforward C programs optimized for efficiency to
generate the large data sets required for useful statistics,
we generated random pure states in the $d^N$ dimensional Hilbert
space, traced over $N-m$ qudits, and analyzed whether the
remaining mixed state, with a $d^m$--dimensional Hilbert space,
had a positive or negative partial transpose (PPT or NPT).
\begin{figure}
    \begin{minipage}{\columnwidth}
	    \begin{center}
	    \resizebox{\columnwidth}{!}{\includegraphics{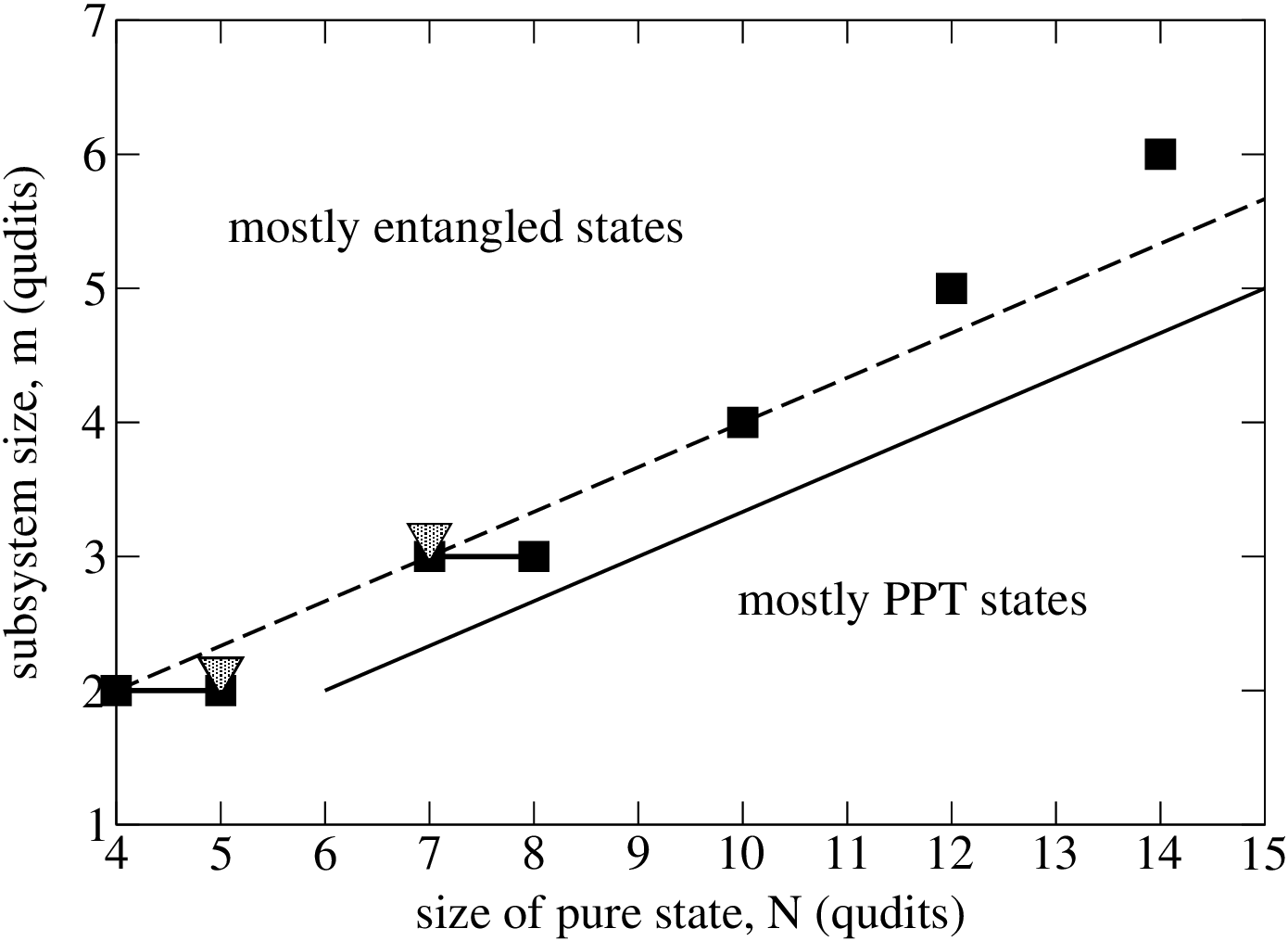}}
	    \end{center}
    	    \caption{Numerical results showing
           	the transition from entangled
           	to PPT subsystems.
           	For each $m$, points are shown for which
	   	$P_{\rm PPT}\in (1\%,99\%)$ for
	   	qubits (\protect\rule[0ex]{1ex}{1ex}) and
	   	qutrits (grey $\bigtriangledown$).
	   	The solid/dashed lines represent
	   	the range in eq. (\ref{eq:twobound}).}
	    \label{fig:zero}
    \end{minipage}
\end{figure}
In fig. \ref{fig:zero} we represent the transition region in the
$N$--$m$ plane for qubits up to $m=6$, $N=15$ and qutrits up to $m=3$, $N=8$,
combining data from \cite{kendon01a} with our new data for
$N=14$ to 15 (qubits) covering the transition for $m=6$,
and qutrits for $N=4$ to 8.
Our new results confirm the trend.  For $m > 4$,
the transition from NPT to PPT for 
a given subsystem size $m$ is completed for $N<3m-2$.
This is due to our estimates not taking into account the full size
of the set of PPT states, being based only on the maximal ball inscribed
in this set.

There is one more feature of this transition that requires comment.
As noted in \cite{kendon01a}, the transition is very sharp, the proportion
of PPT subsystems of a given size $m$ becomes essentially zero to numerical
accuracy with the addition of just a few qudits to the pure state.
It is not obvious \textit{a priori} that the transition should be
sharp.  One might reasonably expect it to be smoother,
with a typical pure state of $N$ qudits containing finite amounts of
$m-$partite entanglement for all $2\le m\le N$.
In order to obtain such a sharp transition,
the proportion of NPT subsystems of given size $m$ must fall exponentially
in the size of the Hilbert space $d^N$ of the pure state.
This is hard to check numerically, but the data we have do
support this, see fig. \ref{fig:pr_npt}.
\begin{figure}
    \begin{minipage}{\columnwidth}
        \begin{center}
            \resizebox{\columnwidth}{!}{\includegraphics{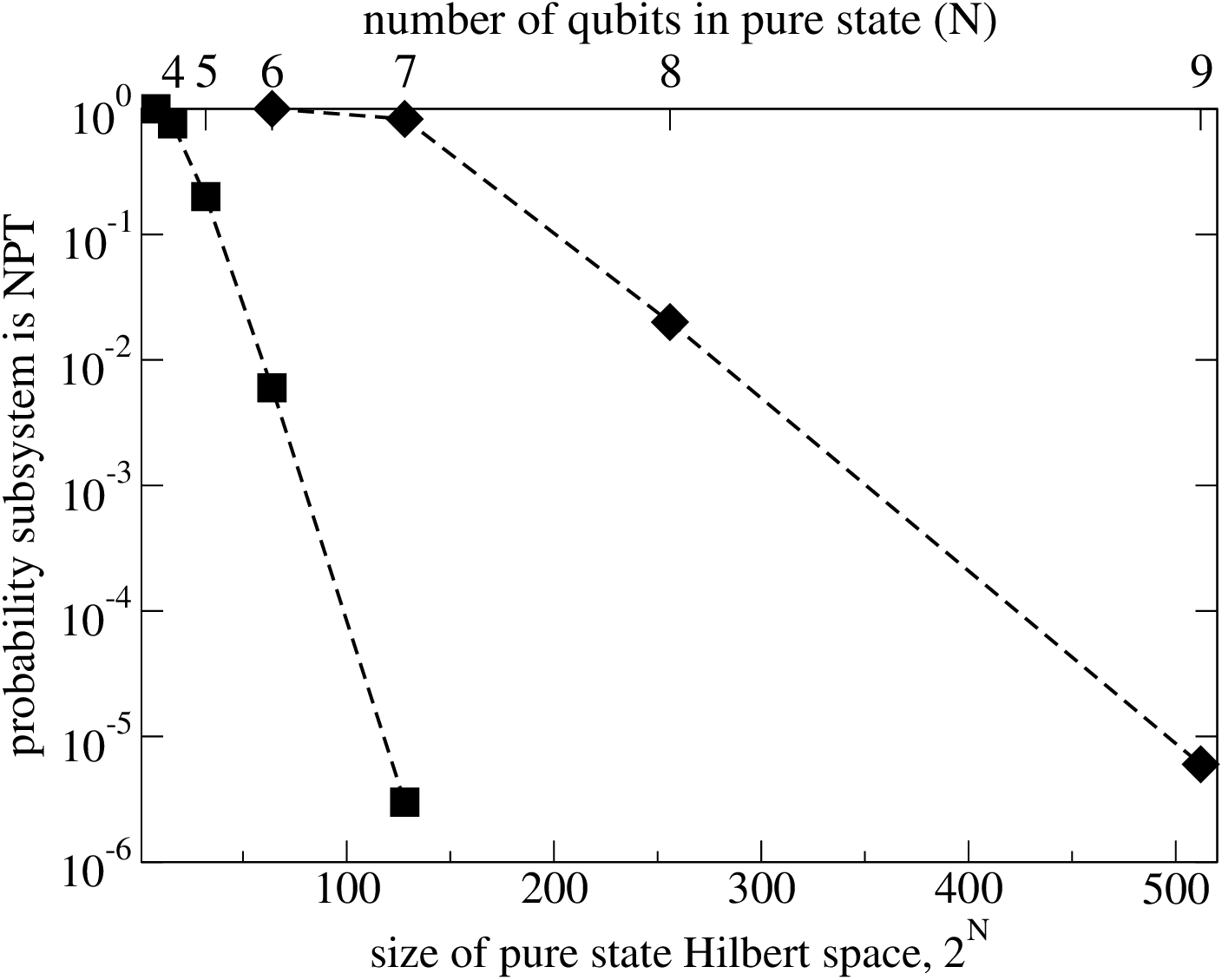}}
        \end{center}
        \caption{Probability of finding that a subsystem of $m$ qudits is
		entangled in random pure states sampled uniformly over the
		Haar measure for qubits for $m=2$ (\protect\rule[0ex]{1ex}{1ex})
		and $m=3$ (filled $\diamond$).}
        \label{fig:pr_npt}
    \end{minipage}
\end{figure}
This sharp transition and exponential fall off allows us to assert that
this bound on the multi-partite nature of entanglement in typical states
of Hilbert space corresponds to the distinction made in \cite{jozsa02a}
between sets of states that can be simulated efficiently classically,
and sets of states that cannot.  It also justifies our statement in the
introduction that exceptions to our bound (\ref{eq:upper})
do not contribute significantly to the average properties of typical
pure states.

\section{Discussion and conclusions}
\label{sec:conc}

We have obtained analytically a lower bound on the size of subsystem
of a typical pure state that can be expected to contain any entanglement
within the subsystem.  For a typical pure state of $N$ qudits, its
subsystems of size $m < N/3$ can be expected to be PPT, i.e. to have
no usable entanglement within them.  We then checked numerically
and confirmed that the transition from NPT to PPT subsystems actually
occurs at significantly larger $m$ nearer to $N/2$, which is the asymptotic
limit as $N\rightarrow\infty$.
Our results emphasize the
importance of multipartite entanglement in quantum information processing.
The types of states necessary for exploring the main bulk
of Hilbert space with quantum computers are highly entangled,
but the entanglement will not be evident if only a small subset of
the qudits are examined.  For a quantum computer operating in a pure state,
at least a third of the qubits would need to be processed to detect the
entanglement present.

\begin{acknowledgments}
It is a pleasure to thank Christof Zalka for useful discussions.
We also thank the referee for helpful comments.
This work was funded in part by
the European project EQUIP(IST-1999-11053) and QUIPROCONE (IST-1999-29064).
VK is funded by the UK Engineering and Physical Sciences
Research Council grant number GR/N2507701.
K{\.Z}  acknowledges the Polish KBN grant no 2P03B-072-19
and is grateful to the Blackett Laboratory for the hospitality
during his stay in London.
\end{acknowledgments}

\appendix

\section{PPT property for mixed states of multi--partite systems}
\label{sec:app_ppt}

In this appendix we prove the following 

\textbf{Proposition}. \textit{Any  mixed state $\rho$  of $m$ qudits 
which  satisfies the condition 
\begin{equation}
 R(\rho)= \frac{1}{{\rm tr} \rho^2}  \geq  d^m-1
\label{eq:ipr-ppt2}
\end{equation}
has the PPT property,
i. e., all its partially transposed matrices
$\rho_m^{T_j}$ are positive.}

We start the proof by invoking an algebraic  

{\bf Lemma}. {\sl Let $A$ be a $n \times n$ non-zero Hermitian matrix
and consider the real number $\alpha:={\rm tr } A /({\rm tr} A^2)^{1/2}$.
If $\alpha \ge \sqrt{n-1}$ then $A\ge 0$};  

the proof of which is given by Mehta,
\cite{mehta89a}, inequality 9.21, p. 217.  
Let us apply it for the analyzed density matrix 
partially transposed in an arbitrary way,
$A=\rho^{T_j}$, which remains a Hermitian matrix of size $n=d^m$.
Any operation of partial transpose $T_j$
does not change the traces, so 
$\text{tr}\rho^{T_j}=\text{tr}\rho=1$ 
and $\text{tr}(\rho^{T_j})^2=\text{tr}\rho^2=1/R$.
Therefore, the coefficient
$\alpha^2=1/{\rm tr} \rho^2$
is just equal to the inverse participation ratio $R$.
If a mixed state $\rho$  satisfies the condition (\ref{eq:ipr-ppt2}),
 the assumption $\alpha \ge \sqrt{n-1}$ is fulfilled. Hence  
matrices $\rho^{T_j}$ are positive for all
possible operations $T_j$ of partial transpose, so the
state $\rho$ is PPT. $\Box$.

It is illuminating to discuss a simple geometric
interpretation of the condition (\ref{eq:ipr-ppt2}), which
defines a certain subset of the convex body of mixed states ${\cal M}$
acting in $n=d^m$ dimensional Hilbert space.
The center of the $n^2-1$ dimensional set ${\cal M}$
 is given by the maximally mixed state 
$\rho_*={\mathbb I}/n$. Introducing the Hilbert--Schmidt metric
\begin{equation}
D_{\rm HS}(\rho_1,\rho_2)=\Bigl({\rm tr}(\rho_1-\rho_2)^2\Bigr)^{1/2}
\label{HSmetr}
\end{equation}
it is not difficult to compute the distance of any state 
$\rho$ with eigenvalues $\{x_i\}$ to the center of ${\cal M}$,
\begin{equation}
D_{HS}^2(\rho,\rho_*) = \sum_{i=1}^n \Bigl( x_i-\frac{1}{n} \Bigr)^2
 =  \frac{1}{R} - \frac{1}{n}.
\label{HSrho}
\end{equation}
Hence the condition $R=$const determines a set of points
equidistant from $\rho_*$.

The boundary of ${\cal M}$ is defined by the condition
det$\rho=0$, so any state belonging to it has
at least one eigenvalue equal to zero.
The state  $\rho_a$ with the spectrum 
$\{\frac{1}{n-1},\frac{1}{n-1},\dots,\frac{1}{n-1},0\}$
 is closest to $\rho_*$
among the states belonging to the boundary of ${\cal M}$.
Its inverse participation ratio, $R(\rho_a)=n-1$,
coincides with the critical value from the constraint 
 (\ref{eq:ipr-ppt2}). Thus all states satisfying it form the 
 maximal ball inscribed into the convex body of mixed states ${\cal M}$
and centered at $\rho_*$.

\section{von Neumann entropy on the boundary of the maximal ball}
\label{sec:app_vNent}

In this appendix we outline the proof of the following

\textbf{Proposition}.
\textit{If the von Neuman entropy of a state $\rho$ is larger than $S_c$,
then $\rho$ belongs to the maximal ball and its inverse participation
ratio $R$ is larger than $n-1$ where $n$ is the dimension of the Hilbert space
of $\rho$ and $S_c$ is given by
\begin{equation}
   S_c=\ln n -\frac{2}{n}\ln 2 +\frac{(n-2)}{n}\ln\left[\frac{n-1}{n-2}\right].
\label{eq:entcrit}
\end{equation}
In the asymptotic limit $n\rightarrow\infty$, $S_C\rightarrow\ln n$, i.e. maximal.}

State $\rho_a$, with the spectrum
\begin{equation}
    \lambda_a=\left\{\frac{1}{n-1}, ...\frac{1}{n-1}, 0\right\},
\end{equation}
belongs to the boundary of the set of mixed states
and to the maximal ball inscribed in it since 
\begin{equation}
    R=\frac{1}{tr(\rho^2)} = n-1,
\end{equation}
see Appendix \ref{sec:app_ppt}.
Its entropy $S(\rho_a)=\ln(n-1)$ is the smallest among all states
located at the surface of the hyperspehere $R=n-1$.
To show this it suffices to perturb the spectrum keeping $R$ constant
and compute the entropy variation.
Due to the symmetry of the problem, the state $\rho_b$
which belongs to this hypersphere
and has the largest entropy, is located opposite to $\rho_a$ with 
respect to the maximally mixed state $\rho_*=\mathbb{I}/n$ and we have
\begin{equation}
   (\rho_b+\rho_a)/2 =\rho_*.
\end{equation}
This gives the spectrum of $\rho_b$
\begin{equation}
    \lambda_b=\left\{\frac{n-2}{n(n-1)} , ... ,\frac{n-2}{n(n-1)}, \frac{2}{n}\right\},
\end{equation}
which evidently belongs to the maximal ball $R=n-1$.

Computing its entropy $S(\rho_b)$ one obtains the 
critical value $S_c$ given by (\ref{eq:entcrit}).
$\Box$.


\bibliography{ent}


\end{document}